\documentclass[11pt]{article}
\usepackage[dvips]{graphicx}
\usepackage{latexsym}
\usepackage{amssymb}

\newcommand{\proof}{\smallskip \noindent {\bf Proof: }}
\newcommand{\M}{\mathcal{M}}
\newcommand{\C}{\mathbb{C}}
\newcommand{\CP}{\mathbb{CP}}

\newcommand{\NN}{\mathbb{N}}
\newcommand{\R}{\mathbb{R}}

\renewcommand{\P}{\mathbb{P}}

\newcommand{\T}{\mathcal{T}}

\renewcommand{\d}{\mathrm{d}}
\newcommand{\SL}{\mathrm{SL}}
\newcommand{\SU}{\mathrm{SU}}
\newcommand{\m}{m}

\def\be{\begin{equation}}
\def\ee{\end{equation}}
 \def\theequation{\thesection.\arabic{equation}}

\def\Sm{\Sigma}

\def\OO{\cal O}
\def\om{\omega}

\def\p{\partial}

\def\ll{\lambda}

\def\OO{{\cal O}}

\newtheorem{theo}{Theorem}[section]
\newtheorem{prop}[theo]{Proposition}
\newtheorem{lemma}[theo]{Lemma}
\newtheorem{defi}[theo]{Definition}

\begin{document}
\title{Solitons and admissible families of rational curves in twistor spaces}

\author{Maciej Dunajski\\
Department of Applied Mathematics and Theoretical Physics, 
University of Cambridge\\
Wilberforce Road,
Cambridge,
CB3 OWA,
UK,\and
Simon Gindikin\\
Department  of Mathematics, Hill Center, Rutgers University,\\
110 Frelinghysen Road, Piscataway, NJ 08854-8019, USA,
\and
Lionel J. Mason\\ 
Mathematical Institute,
University of  Oxford \\
24-29 St Giles, Oxford OX1 3LB, UK.
}
\date{October 1, 2004}

\maketitle

\abstract{It is well known that twistor constructions can be used to
  analyse and to obtain solutions to a wide class of integrable
  systems.  In this article we express the standard twistor
  constructions in terms of the concept of an admissible family of
  rational curves in certain twistor spaces.  Examples of of such
  families can be obtained as subfamilies of a simple family of
  rational curves using standard operations of algebraic geometry. By
  examination of several examples, we give evidence that this
  construction is the basis of the construction of many of the most
  important solitonic and algebraic solutions to various integrable
  differential equations of mathematical physics.  This is presented
  as evidence for a principal that, in some sense, all soliton-like
  solutions should be constructable in this way.}

\section{Introduction}
Penrose's nonlinear graviton construction \cite{Pe76} realizes any
4-dimensional conformal manifold with anti-self-dual Weyl tensor as a
family of rational curves in a complex 3-manifold, $\T $, known as the
twistor space.  There are many generalizations, most notably the
extension by Ward to the anti-self-dual Yang-Mills equations
\cite{Wa77}, in which any anti-self-dual Yang-Mills field can be
reconstructed directly from a family of rational curves in the total
space of a holomorphic vector bundle over a region in twistor
space.\footnote{This construction is usually viewed in the context of
  finding trivialisations of the holomorphic vector bundle over the
  lines in twistor space, but this, by Liouville's theorem, is
  equivalent to lifting the the lines in twistor space to the total
  space of the holomorphic vector bundle.}  Similar constructions
apply to a large family of integrable systems both by considering
families of rational curves in larger complex manifolds, or by
symmetry reduction of these two basic twistor correspondences, see
\cite{MW96} for a catalogue of such reductions and full details of the
twistor correspondences.

In effect, the task of solving these equations reduces to that of the
construction of such families of rational curves.  A strategy proposed
by one of us \cite{G82,G85,G90} is to consider {\em Admissible
  families} of rational curves.  An admissible family of curves is a
local family of curves that admit an extension in some larger complex
manifold to a family of compact rational curves that is complete (in
the sense of Kodaira) family of rational curves.  To construct such
families, one can consider a simple space such as the total space of
some bundle $\OO(k_1)\oplus\OO(k_2)\oplus\cdots \oplus\OO(k_r)
\rightarrow \CP^1$, $k_1 \geq k_2\cdots \geq k_r\geq 0$ that has too
many rational curves; here the space of sections has dimension
$\sum_{i=1}^r (k_i+1)$.  One can then restrict the family in such a
way that the restricted family is an admissible family with the
appropriate dimension and so can be regarded in its own right as a
moduli space of rational curves in some (different) twistor space.
There is a theorem due to Bernstein \& Gindikin \cite {G84, BG03} to
the effect that admissible families can only be obtained from some
large family by requiring intersection with or tangency to some family
of submanifolds. This, in turn, corresponds to the operation of taking
a blowup of or branched cover over the original twistor space at the
chosen the submanifolds. These algebraic geometrical modifications of
the twistor space can be thought of as finite non-linear deformations
of the twistor space in Penrose's nonlinear graviton construction.

Such a procedure will not give rise to the general solution as these
methods use data on subspaces of codimension at least 1 to that
required for the general solution.  Thus these solutions are special.
In this paper, our aim is to show that many of the most important
soliton-like solutions in mathematical physics arise from this
construction.  In particular, standard procedures for obtaining
special solutions by introducing a hierarchy and imposing hidden
symmetries fit naturally into this construction.  We present this as
evidence for a principal that all interesting soliton-like solutions
arise in this way.

In Section 2 we outline the basic methodology associated to families
of rational curves. In section 3 we explain how the imposition of
restriction on a simple but large admissable family of rational curves
arises naturally in situation in which hierarchies are considered and
higher symmetries imposed.  In the subsequent sections we show how various
standard global solutions (solutions arising from the Ward ansatze
such as instantons and monopoles in $\R^3$, Korteweg de Vries and
non-linear Schrodinger solitons, and ALE gravitational instantons) fit
into this framework.  For the non-linear graviton construction, the
construction of families of rational curves is connected only with the
conformal part of the problem. In particular, the problem of
constructing a self-dual 4-metric on $\M$ is equivalent to
constructing a pair of 1-forms $\varphi(t), \psi(t)$ depending on a
rational parameter $t$ such that the 2-form $F(t)=\varphi(t) \wedge
\psi(t)$ is quadratic in $t$ \cite {G82}. The conformal part of this
problem is the integrability of the kernel distribution if
$\varphi(t)$ and $\psi(t)$ for all $t$.  This weaker problem can be
reformulated as a problem of finding an $\OO(1)\oplus\OO(1)$-family of
rational curves. As a result we need to modify the
tangency-intersection construction in such a way that we obtain
solution to the complete problem, not just its conformal part.  Thus,
in Section 5.2 on the construction of ALE solutions, we work with a
special type of tangency-intersection condition connected with the
lifting of rational curves on certain branched coverings.  In the
subsequent sections, we demonstrate that classes of three-dimensional
Einstein--Weyl spaces, and ODEs satisfying a set of over-determined
W\"unschmann--type constraints can be reconstructed from admissible
families.  The basic facts about bundles over $\CP^1$ and the
deformation theory are summarised in the Appendix.

We are grateful to the anonymous referees for their comments that led
to important improvements to this article.
MD was partly supported by NATO grant PST.CLG.978984.

\section{Admissible families of rational curves.}
In twistor theory, the data of a solution to an integrable equation is
encoded into the complex structure of the twistor space $\T$.  The
reconstruction of the solution from the complex structure on $\T$
reduces to the construction of families of rational curves in $\T$.
The parameter space $\M $ of the family is either a space-time or some
related space (i.e., the total space of a Yang-Mills bundle over
space-time) and the full solution to the integrable equation can be
obtained directly from the correspondence with the rational curves in
$\T$.  In this article, we study such families of rational curves
locally in $\T$, and the global condition that the curves extend to
global rational curves in $\T$ is replaced by the condition that the
family be admissible as defined below.

A family of rational curves in $\T$ is complete if it contains all
small deformations in the family.  We will give a {\em local } (in
$\T$) characterization of complete families of rational curves.  Since
we are working locally, this characterization will be birationally
invariant.  Roughly speaking, this will mean that, perhaps after some
extension and modification of $\T$, it will be the family of {\em all}
rational curves of some fixed topological type in some manifold
$\widetilde \T$.  However, after another birational map, it may cease
to be a maximal in the new manifold (and in general, it will not be
maximal in $\T$). We will work in the complex holomorphic category
throughout.

Initially, we will define families of local rational curves (which may
not necessarily extend to become global). Let $\T $ be a
complex manifold of dimension $r+1$ in which a family of curves are
embedded, and let $\M $ be a manifold of  dimension $n$,
parametrising the curves with $\m\in \M $ corresponding to the curve
$C_m$ in $\T $. Our considerations are local in the first instance,
and we will assume that everything is in general position.

For $z\in \T $, let $Z$ be the dual submanifold in $\M$, which
parametrises the curves $C$ that pass through $z$. Such a $Z$ will be
called an $\alpha$-surface.  Inside the tangent space $T_{\m}\M$ at
$\m\in \M$, we define an $\alpha$--plane to be a tangent plane
$\alpha(z)$ to some $Z$, for which $z\in C_m$.  Let $v_\m $ be the
union of these $\alpha$-planes in $T_\m$.

\begin{defi}
We will say that $v_\m$ satisfies the {\em locally rational}
condition (LR) if it can be presented as the union of 2-dimensional
flat cones, i.e., cones sitting inside 2-dimensional subspaces of
$T_\m\M$.  These 2-planes, called $\beta$-planes, will be denoted
$\beta(s)$ and we will assume that they intersect only at 0 and are
transversal to the planes $\alpha(x)$. Let $V_\m$ be the union of
these 2-planes, clearly $v_m \subset V_m$.
\end{defi}

\begin{lemma}
If the curves do in fact admit an extension so that they form a
complete family of global rational curves, then the LR condition is
satisfied.
\end{lemma}

\proof In the case where the curves are global, the normal bundle
$NC_m$ of a rational curve $C_m$ in $\T $ will be
$\oplus_{i=1}^{r}\OO(k_i)$ for some integers $k_i$ and we will assume
that these integers are constant in the family.  (Here $\OO(k)$
denotes the line bundle of Chern class $k$ on $\CP^1$.)  By Kodaira's
theorem, $T_\m\M=\Gamma(\CP^1_\m, N)$.  If we express $\CP^1$ as the
projectivisation $\P S^*$ of a 2-dimensional complex vector space
$S^*$, we obtain $T_\m\M=\oplus_{i=1}^{n-1}\odot^{k_i}S$ (and so
$n=\sum_i (k_i+1)$).

Let $\pi \in S$ then the general tangent vector to the $\alpha$-plane
corresponding to the point where $\pi$ vanishes on $\P S^*$ will lie
in the image of the vector space
$W=\oplus_{i=1}^r\odot^{k_i-1}S$ embedded into $T_\m\M$ by
symmetrization of each summand with $\pi$.  Hence, each element of
$\P(W)$ corresponds to the two--plane in $T_\m\M$ spanned by the
image of symmetrizing each summand with $S$. $\Box$

\smallskip

The structure of the incidence cones $v_\m \subset T_\m\M$
supplies the curves $C_m$ with the canonical projective structure
of rational curves, induced by the projectivisation of one of the
2-planes $\beta(s)$.  This structure is independent of the choice
of $s$. If we consider curves $C_m$ without parameterization, we
can identify vectors in $T_\m \M$ with sections of the normal
bundles $NC_m$. For manifolds of parameterized curves we can
interpret them as sections of tangent bundles $TC_m$.

Even in the general local case, the cones $V_\m$ must be linearly
equivalent to cones for the case when $\M$ is the manifold of
sections of a vector-bundle $\mathcal O(k_1)\oplus \cdots \oplus
\mathcal O(k_r)$,  $k_1\ge \cdots \ge k_r\ge 0$.  These families of
curves play the role of the flat model in this geometry.  In the
generic situation, $V_\m$ for different $\m\in \M$ are linearly
equivalent and correspond to some choice of $ k_1\ge \cdots \ge k_r\ge 0$.

This gives the manifold $\M$ a generalized conformal structure $\{
V_\m\}$.  It is natural to investigate when this structure uniquely
defines the manifold $\T $ with the curves $C_\m$. The first step is to
understand when the $\alpha$ and $\beta$ planes can be uniquely
reconstructed. It turns out that this is possible if $V_m\neq\C^p$ for
some $p$, which is equivalent to the condition $k_2>0$.  Then the
$\alpha$-surfaces, $Z$
are defined as integral submanifolds whose tangent planes are
$\alpha$-planes at every point.  It turns out that, with these
conditions, the $Z$, if they exist, are unique (although existence
is not automatic).  Then we can construct $\T $ as the manifold of
integral submanifolds of such $Z$, and the curves $C_m\subset \T $
defined as the sets of solutions passing through $\m$.

For the definition of $Z$, in general it is necessary
to produce an equation of second order, or to consider over $\M$ the
fibering whose fiber over $\m \in \M$ parameterizes the set of
$\alpha(z)\subset T_\m$ and a Frobenius distribution whose
projection at an $\alpha$-plane $Z$ to $\M$ consists of the
$\alpha$-plane $Z$. Under the condition $k_2>0$, this
Frobenius distribution can be uniquely reconstructed from
the integrability conditions.

\begin{prop}[Generalized Desargues theorem \cite{BG03,G90}.] If the $V_\m$ are
  subspaces of $T_m\M$ which are different from $\C^p$, $p\leq m$
  (or equivalently $k_2>0$), then the field $V_\m$ determines
  at most one family of $\alpha$-surfaces parametrized by a manifold
  $\widetilde \T$.  If $V_m\subset T_m\M$ arose from a family of
  curves $C_m$ in a manifold $\T$, then the manifold $\T $ with curves
  $C_m$ can be reconstructed uniquely as a subset of $\widetilde\T$.
  All curves $C_m$ admit a canonical extension up to global rational
  curves (after the appropriate extension of the manifold $\T
  $ to $\widetilde\T$).
\end{prop}

In the general case (when the condition of the above proposition is
not satisfied) the extendibility of the curves $C$ to global rational
curves requires an algebraic condition on the family of
$\alpha$-planes. The simplest way to specify this is to use the (LR)
condition as follows.  We will say that a family of curves $C$
satisfies the (R) condition if there exists an (LR)-family $\widetilde
C$ parametrized by $\widetilde \M$ consisting of the family of all
curves $C$ together with projective parametrizations that are
compatible with a local projective structures on them corresponding to
a decomposition of $V_m$ into $\beta$-planes. This gives an
$n+3$-dimensional family $\widetilde \M$ of curves fibred over $\M$
with fibres $\mathrm{PSL}(2,\C)$.  An (LR)-structure on $\widetilde
\M$ is what we define to be an $(R)$-structure on $\M$. This
$(LR)$-structure on $\widetilde \M $ defines a canonical extension of
curves $C_m$ in $\widetilde \T$ to global rational curves in a larger
manifold. The essential point here is the existence of a canonical
decomposition of cones $V_\m$ into $\alpha$-subspaces transversal to
$\beta$-subspaces.

Let us remark that when $n=r+1=2$ the condition (R) is equivalent
to the Cartan condition: curves $U(m)$ are defined by a differential
equation of 2-nd order which is a polynomial of degree 3 in the first
derivative, i.e., they define a projective structure.

A principal result of the theory of families of rational curves
concerns `admissible' subfamilies:
\begin{defi}
Given a (perhaps local) family of rational curves  with the property
(R), a subfamily will be said to be admissible if the property (R) is
induced on it.
\end{defi}
It emerges that only very special subfamilies have this a
property. Let us start from the generic case.

\begin{prop}\label{admissibility}
  Let $\M$ be a family of (global) rational curves on $\T $.  Let
  $T_1,..., T_p$ be submanifolds in $\T $ of codimensions more then 1
  and $S_1,...,S_q$ have codimension 1 and $s_1, ... ,s_q$ are natural
  numbers. Let $\M (T,S,s)$ be the subfamily of curves $C_m$ which
  intersect all $T_i$ and have at their intersection with each $S_j$
  tangency of order $s_j$. Then this subfamily is admissible and a
  generic admissible subfamily can be represent in such a form.

To obtain all admissible subfamilies, not just the generic ones, we
need to replace the conditions of intersection with the $T_j$ by the
condition that we take those curves that admit a lift to a tower
$\widetilde \T $ of blow ups of $\T $. Then we consider the hypersurfaces
$S_j$ to be in $\tilde \T $ and add the condition of tangency of order
$s_j$.
\end{prop}

There are two known proofs of this theorem: one use algebraic
geometrical methods \cite {BG03} and the other one uses geometrical
methods of nonlinear differential equations \cite {G84}.  The proof
of this theorem splits into several steps: the condition of
admissibility of a subfamily $\widetilde \M \subset \M $ can be
written down as an explicit nonlinear differential equation. It
turns out that this equation can be integrated by a generalized
Hamilton-Jacobi method with multidimensional bicharacteristics.

\section{Hierarchies for the anti-self-duality equations and
  invariance under hidden symmetries}

A key feature of soliton solutions to, for example, the KdV equations
is that they arise as solutions that are invariant under one or more
hidden symmetries. A hidden symmetry is usually understood in the
context of hierarchies associated to the equation.  

A hierarchy associated to an integrable system is an overdetermined
system of (completely integrable) partial differential equations on a
higher-dimensional space, usually the cartesian product of the
space-time for the original integrable system with a space of higher
times.  The hierarchy equations restrict to give the original
completely integrable system on each leaf of the foliation on which
the `higher time' variables are constant.  The flows along higher time
variables are know as hidden symmetries since they evolve solutions to
the original equations into different solutions to the original equations.

The connection with the above theory is that the solution to the
hierarchy will arise from some admissible family of rational curves,
and each solution to the original system at fixed values of the higher
times arises from an admissible subfamily.  Proposition
\ref{admissibility} implies that such admissible subfamilies must be
obtained by intersection and tangency.

A standard strategy for obtaining soliton solutions is to require that
a solution can be embedded into a solution to the hierarchy that is
invariant under one or more higher flows.  More generally, one can
require that the solution to the hierarchy admits one or more
symmetries that are not necessarily symmetries of the original system.
Thus the solution to the hierarchy can be taken to arise from a simple
admissible family, and the solitonic solution to the original system can be
obtained by intersection and tangency as described above.

The twistor correspondences generalize straightforwardly to the
hierarchy.  In the case of the Bogomolny equations, the twistor space
corresponding to a solution is the total space of a holomorphic vector
bundle over `minitwistor space', $\OO(2)$ (the total space of the line
bundle of Chern class 2 over the Riemann sphere, $\CP^1$).  In
\cite{MS92} it was shown that the $\SU(N)$ Bogomolny equations embed
into a hierarchy, referred to as the Bogomolny hierarchy, for which
the twistor space is the total space of a holomorphic vector bundle
$\OO(n)$ for some (arbitrarily large) $n>2$.  The rational curves in
this space have normal bundle that is a direct sum of the trivial
$\C^N$ bundle with $\OO(n)$.  This yields the standard hierarchies for
the KdV and the nonlinear Schrodinger equations under symmetry
reduction.  This was extended in \cite{MW96} to a correspondence for
hierarchies for the $\SU(N)$ ASDYM equations, in which a solution to
the hierarchy corresponds to a holomomorphic vector bundle over the
total space of $\OO(n)\oplus\OO(n)\rightarrow \CP^1$ so that the
rational curves have normal bundle given by the direct sum of $\C^N$
with $\OO(n)\oplus\OO(n)$.  In \cite{DM00,DM03} the twistor
correspondences were extended to give a hierarchy for the
hyper-K\"ahler equations in which the twistor space is as usual, a
3-dimensional complex manifold fibred over $\CP^1$ and admits a
Poisson structure on the fibres, but now the family of rational curves
has normal bundle $\OO(n)\oplus\OO(n)$. 

It is natural, therefore, to regard the geometry arising on
the moduli space of rational curves with some arbitrary, but fixed
normal bundle, $\OO(k_1)\oplus\OO(k_2)\oplus\cdots
\oplus\OO(k_r) \rightarrow \CP^1$, $k_1 \geq k_2\cdots \geq k_r\geq 0$,
as the most general hierarchy associated to equations that admit a
twistor correspondence. This can be though of as a set of differential
equations implied by the (LR)-condition on the generalised conformal
structure formed by the family of incidence cones $V_m$ in each
$T_m\M$.

A solution to some version of the anti-self-duality equations will
then extend to a hierarchy if it can be realised as arising from an
admissible subfamily of the family of rational curves associated to
that hierarchy.  By proposition \ref{admissibility}, if one is just
given the solution to the hierarchy, admissible subfamilies are found
by requiring intersection or tangency to submanifolds and all
admissible subfamilies arise in this way.

A key application of hierarchies is to perform symmetry reduction, but
with respect to a `hidden' symmetry and in practice soliton solutions
often arise in this way.  This will mean that we will consider a
solution that can be embedded into a hierarchy that admits at least
one explicit symmetry.  The hierarchy can admit many symmetries
without the original solution admitting any at all.

If one wishes to look for such solutions, one can consider simple
solutions to the hierarchy, perhaps even trivial ones corresponding to
a constant conformal structure or flat connection, but then find
non-trivial solutions by using intersection and tangency to find
a non-trivial admissible subfamily.

\section{Examples}
We have seen then that solutions of many problems of mathematical
physics that can be integrated by the inverse problem method require
the construction of some family of rational curves. Using Proposition
\ref{admissibility}
it is possible to produce such families as (admissible)
subfamilies of some simple families of rational curves depending of a
larger number of parameters, for example, the families of sections of
a vector bundle on the projective line. Let us remember that these
families play the role of flat objects in this geometry and we are
interested in solutions that admit embedding in flat solutions of a
bigger dimension. Of course, not all local solutions can be produced
such a way (such solutions depend on fewer functional parameters and
will be partly algebraic) but we can expect that some `good' global
solutions can be included in this construction. They are in a sense
quasi soliton solutions. We show here that it is indeed the case for
several important problems: we will find intersection-tangency
conditions in their solutions.

We believe that it is realistic to build Ansatze for solutions for a
number of problems starting from these ideas.

\subsection{The Ward ansatze and intersection conditions}
In the case of the Ward construction for solutions to the ASD
Yang-Mills equations, the twistor space $\T$ is the total space of a
holomorphic vector bundle $E\rightarrow U$ where $U$ is some region in
$\CP^3$ and $E$ is assumed to be trivial on each real line in $U$
(i.e., on each line that is invariant under some anti-holomorphic
conjugation $\sigma:U\rightarrow U$).
Usually, the key step in the construction of a solution is the task of
finding an explicit trivialisation of $E$ over the (real) lines in $U$.  This
is equivalent to finding the rational curves in $E$ that are sections
of $E$ over each line in $U$.

The process of finding a trivialisation of $E$ over a line in $U$
requires, in the Cech description of the bundle, the solution to a
Riemann-Hilbert problem and this is difficult to find explicitly.
However, it can be done explicitly in the case where the patching
matrix is upper triangular and this is the Ward ansatze.  When $E$
has rank two with structure group $\SL(2,\C)$ (which we will assume
from hereon), this can be expressed as the requirement that $E$
contains a line subbundle $L^*$ and sits in a short exact sequence:
$$
0\longrightarrow L^*\longrightarrow E \longrightarrow L
\longrightarrow 0\, .
$$ 
The line bundle $L$ must have non-negative degree $k \geq 0$ on
each line if $E$ is to be trivial on each line in $U$ (a positive
degree line subbundle $L^*$ would contradict triviality, but not a
negative degree one).

The assumption that $E$ admits such a line bundle (or that the
patching matrix can be expressed in upper triangular form) is known
as the Ward ansatze and has been very fruitful in constructing
solutions to the anti-self-dual Yang-Mills equations and its
reductions, see Ward (1981).  In particular, all instanton solutions
can be obtained in this way \cite{AW77}, all monopoles \cite{Hi82},
and solitons for the non-linear Schrodinger equations \cite{MS92}.
The following discussion is a paraphrase of Hitchin's discussion of
monopoles, \cite{Hi82}.

When the solution to the anti-self-dual Yang-Mills equations on
space-time has gauge group $\SU(2)$ or $\SU(1,1)$ on a real slice, the reality
conditions imply that the anti-holomorphic involution
$\sigma:U\rightarrow U$ lifts to give give an isomorphism between
$\overline {\sigma^*E}$ and $E$ (see Atiyah 1979 or Mason \& Woodhouse
1996 for a full discussion).

We assume that $\hat L^*:=\overline{\sigma^*L^*}$ is generically
a linearly independent line subbundle to $L^*$.  In this
case the intersection/tangency ideas of the previous section can be
brought into play since we have a map
$$
\rho:E\hookrightarrow L\oplus
\hat L\, ,
$$
and this will be a fibrewise vector space isomorphism except on the
codimension-1 set $T\subset U$ on which $L^*$ and $\hat L^*$ coincide
as line subbundles of $E$.  Denote the image of $\rho$ over $T$ by
$\widetilde T$.  (Since on $T$, $\rho$ is onto both $L$ and $\hat L$
separately, $\widetilde V$ is the graph of an invertible map from $L$
to $\hat L$, or alternatively a trivialisation $e$ of $L\otimes \hat
L^*$.)

The data of $L$, $\hat L$ and $\widetilde V$ is sufficient to
reconstruct the original solution.  Sections of $E$ over lines in $U$
correspond precisely to sections of $L\oplus \hat L$ that pass through
$\widetilde T$.  Since the line bundles $L$ and $\hat L$ have degree
$k$, there will be $2k+2 $ sections of $L\oplus \hat L$ over each line
in $U$.  The submanifold $T$ must therefore have degree $2k$ so that
there are $2k$ conditions on the $2k+2$ sections, reducing the number
of sections of $E$ to $2$ as required by triviality on lines.

We therefore have

\begin{prop}
  Suppose $E\rightarrow U$ is a rank two holomorphic vector bundle
  such that $\overline{\sigma^* E}=E$ and there is a line subbundle
  $L^*\subset E$ such that $\hat L^*:=\overline{\sigma^* L^*}$ is
  generically linearly independent from $L^*$.  Then the admissible
  family of rational curves consisting of sections of $E$ over lines
  in $U$ is equivalent to the admissible subfamily of sections of
  $L\oplus \hat L$ over lines in $U$ that intersect the codimension-2
  subset $\widetilde T$ of $L\oplus \hat L$ as defined above.
\end{prop}

\subsection{The monopole solutions}
Monopoles are solutions to the ASD Yang-Mills equations on $\R^4$ with
a single translation symmetry and with a finite energy condition.  In
\cite{Hi82} Hitchin showed that monopoles were, via the Ward
correspondence, in a correspondence with holomorphic vector bundles on
$T\CP^1$ that are constructed from a certain `spectral curve' $T$
in $T\CP^1$.  The construction involves the use of homogeneous line
bundles $L(n)\rightarrow T\CP^1$ that can be described as
follows.  Introduce affine coordinate $\ll$ on $\CP^1$ and
corresponding fibre coordinate $\eta $ on $T\CP^1$ so that
$(\eta,\ll) \leftrightarrow \eta\p/\p\ll$.  We can define $L(n)$
by the transition function $\ll^{-n}\exp(\mu/\ll)$
with respect to the open covering $U_0=\{ \ll\neq \infty\}$ and
$U_\infty=\{\ll\neq 0\}$.  ($\OO(n)$ is the pullback of the $-n$th
power of the tautological bundle on $\CP^1$ and $L(0)$ is the exponential
of the class obtained by pulling back the `unit element' of
$H^1(\CP^1,T^*\CP^1)$ and contracting it with the vector
$\eta\p/\p\zeta$, and $L(n)=L\otimes\OO(n)$.)

Hitchin shows that if $E\rightarrow T\CP^1$ is the Ward transform of a
monopole solution, then we can express $E$ as an extension in two
different ways
$$
0\rightarrow L(-k) \stackrel{\iota_+}\rightarrow E
\stackrel{\sigma_+}\rightarrow L^*(k)\rightarrow 0\, , \qquad
0\rightarrow L^*(-k) \stackrel{\iota_-}\rightarrow E
\stackrel{\sigma_-}\rightarrow L(k)\rightarrow 0
$$
where $L^*(k)=(L(0))^*\otimes\OO(k)$.  The map $\sigma_+\iota_-$
determines a section $\psi$ of $\OO(2k)$.  The zero set of $\psi$ is
the spectral curve $T$.  Such spectral curves determine $E$ and are
characterised by the conditions that (i) they are compact, (ii)
invariant under the real structure $\tau:(\eta,\ll)\rightarrow
(-\bar\eta/\bar\ll^2, -1/\bar\ll)$ and (iii) the line bundle
$L(0)^2$ is trivial on restriction to $T$.

This construction fits into the previous discussion by virtue of the
fact that, in order to reverse the Ward construction to reconstruct
the solution to the Bogomoln'yi equations, one must first find a
holomorphic trvialisation for $E$ over each section
$\sigma_x:\CP^1\rightarrow T\CP^1$ that is invariant under the real
structure $\tau$, for $x\in\R^3$.  This is equivalent to finding the
appropriate family of holomorphically embedded rational curves in the
total space of $E$ that cover $\sigma_x$.  By projection in each of
the two short exact sequences above, these curves are a subset of
those in $\tilde E=L(k)\oplus L^*(k)$ over $\sigma_x$.  They can be
characterised as the curves that intersect the codimension-2 subset of
$\tilde E$ consisting of the the line subbundle $\tilde T$ of $\tilde E|_T$
defined by the the trivialisation of $L(0)^2$ over $T$.

It is worth noting that it is straightforward to find the
trivialisation of $\tilde E$ over each $\sigma_x$.  This is set out in
\cite{Hi82} for $L(0)$ and one can simply multiply that section by a
pair of polynomials of degree $k$ in $\zeta$ leading to $2k+1$
sections.  Given the spectral curve, $T$ and the trivialisation $e$ of
$L^2(0)$ over $T$, the incidence condition with $\tilde T$ is $2k$
conditions on the $2k+1$ unknowns and leads to the desired
2-dimensional space of sections of $E$.

The implementation of this as a strategy for writing down exact
solutions of monopole is hard.  The principal difficulty is in
choosing the spectral curve $S$ (the simplest examples, the
axisymmetric solutions, are known, but in general the triviality of
$L(0)^2$ on $T$ is a hard condition to impose explicitly).  Even then,
one must solve for the trivialisation of $L(0)^2$ on $T$ and then find
the intersection points of $T$ with a generic $\sigma_x$ in order to
impose the incidence condition with $\tilde T$.

\subsection{Korteweg de Vries and Non-linear Schrodinger solitons}
In \cite{MS92} it is shown that NLS solitons can be obtained from the
Ward ansatze also, and so can be obtained from imposing intersection
conditions on some simple admissible family.  Here we can see that the
construction can be made very explicit.

The KdV and NLS equations arise as symmetry reductions of the
$\SL(2,\C)$ Bogomolny equations in 2+1 signature under a null
translation.  The corresponding holomorphic vector bundles over
regions in $\OO(2)$ have rank two and admit a lift of the symmetry
$K=\p_\mu$ where $(\lambda \mu)$ are coordinates on $\OO(2)$, with
$\lambda $ being an affine coordinate on the base $\CP^1$ and $\mu$
the fibre coordinate on $\OO(2)$.  Since $\OO(2)=T\CP^1$, the choice of
$\lambda$ determines the trivialisation in which $\mu$ corresponds to
the tangent vector $\mu\p/\p \lambda$.

Over a neighbourhood of $\lambda=\infty$, we use coordinates
$(\lambda',\mu')= (1/\lambda,\mu/\lambda^2)$.  In these coordinates
$V=\lambda'^2\p/\p \mu'$ and so $K$ fixes the fibre at $\infty$ to 2nd
order.  The symmetry reductions to the KdV and NLS equations are
distinguished by the action of the lift $\widetilde K$ of $K$ to the
bundle $E$ at $\lambda=\infty$ to second order: we have $\widetilde
K=\Lambda+O(1/\lambda^2)$ where
$$
\mbox{For NLS} \qquad \Lambda=\begin{pmatrix}{1&0\cr 
  0&-1}\end{pmatrix}\, , \qquad \mbox{ and for KdV} \qquad
\Lambda=\begin{pmatrix}{0&1\cr 1/\lambda&0}\end{pmatrix}\, .
$$ 
The trivial solutions correspond to bundles $E_0$ with $\widetilde K$
given as above on a neighbourhood of $\lambda=\infty$ and patching
function $\exp(\mu\Lambda)$ to a trivialisation that extends over
$\lambda\neq\infty$ in which $\widetilde K$ has trivial lift.

The solitons can be obtained from the trivial solutions by considering
first $E_0(k)=\OO(k)\otimes E_0$.  This bundle has $2k+2$ sections
over each conic in $\OO(2)$.  We choose an admissible subfamily by
choosing first $k$ points $\{\lambda_1,\cdots , \lambda_k\}$ in the
upper half plane in $\C$, and then an invariant section $\gamma_i$ of
$\P(E_0^*)$ over each fibre $\lambda=\lambda_i$ of $\OO(2)$.  We
require that $\gamma_i$ does not lie in an eigenspace for $\Lambda$. 
This data determines $2k$ codimension-2 submanifolds of $E_0(k)$, the
kernel of $\gamma_i$ over  $\lambda=\lambda_i$ and the kernel of
$\bar\gamma_i$ over $\lambda=\bar\lambda_i$.  (Note that
$\bar\gamma_i$ needs to be interpreted appropriately according to the
reality condition that we wish the final solution to satisfy).

In order to see explicitly that this yields the standard formulae for
the appropriate admissible subfamily of the space of sections of
$E_0(k)$ over a given conic $C$ in $\OO(2)$, we first represent the
sections of $E_0(k)$ over $C$ as sections of $E_0$ that have simple
poles at each $\lambda_i$.  The residues must lie in the kernel of the
corresponding $\gamma_i$ and $k$ of the remaining $k+2$ coefficients
are fixed by the $k$ conditions of lying in the kernels of the
$\bar\gamma_i$ at $\bar\lambda_i$.  See example 9.3.3 and section 12.4
from \cite{MW96} for further details of such solutions.

\subsection{The ALE solutions}
\label{ale}

In the previous subsections we have seen that see that we can express
many of the most familiar soliton/instanton solutions in terms of
subfamilies of some simple admissible family by imposing intersection
conditions.  In this subsection we see that the the ALE hyper-Kahler
solutions (gravitational instantons) can be expressed as an admissible
subfamily of a simple family by imposing tangency conditions.

Hyper-K\"ahler manifolds $({\cal M}, g)$ that have the topology of
$\R^4$ at infinity, and approach the flat Euclidean metric $\eta=\d
{x_1}^2+...+\d {x_4}^2$ sufficiently fast, in the sense that
\[
g_{ab}=\eta_{ab}+O(r^{-4}),\qquad (\p_a)^p(g_{bc})=O(r^{-4-p}),\qquad
r^2=x_1^2+...+x_4^2
\]
have to be flat. A weaker asymptotic condition one can impose is that
$g$ should be asymptotically {\em locally} Euclidean (ALE).

The ALE spaces are non-compact, complete hyper-K\"ahler
manifolds which satisfy the above condition only
locally for $r\rightarrow \infty$. Globally the neighbourhood of
infinity must look like $S^3/\Gamma\times \R$, where $\Gamma$ is a
finite
group of isometries acting freely on $S^3$ (a Kleinian group).
These manifolds belong to the class of {\em gravitational instantons}
because their curvature is localised in a `finite region` of a space-time.

Finite subgroups of $\Gamma\subset SU(2)$ correspond Platonic solids
in $\R^3$.  They are the cyclic groups, and the binary dihedral,
tetrahedral, octahedral and icosahedral groups (one can think about
the last three as M\"obius transformations of $S^2=\CP^1$ which leave
the points corresponding to vertices of a given Platonic solid fixed).
Each of them can be related to a Dynkin diagram of a simple Lie
algebra.  All Kleinian groups act on $\C^2$, and the `infinity`
$S^3\subset\C^2$. Let $(z_1, z_2)\in\C^2$.  For each $\Gamma$ there
exist three invariants $x, y, z$ which are polynomials in $(z_1, z_2)$
invariant under $\Gamma$.  These invariants satisfy some algebraic
relations which we list below: {\small
\begin{center}
\begin{tabular}{p{1cm}|lll}
\multicolumn{4}{c}{}\\
&{\bf Group} & {\bf Dynkin diagram} & {\bf Relation} $F_{\Gamma}(x,y,z)=0$\\
&cyclic            &$A_k$& $xy-z^k=0$\\
&dihedral          &$D_{k-1}$&$ x^2+y^2z+z^{k}=0$\\
&tetrahedral       &$E_6$& $x^2+y^3+z^4=0$\\
&octahedral        &$E_7$& $x^2+y^3+yz^3=0$\\
&icosahedral       &$E_8$& $x^2+y^3+z^5=0$
\end{tabular}
\end{center}
\small}
In each case
\[
\C^2/\Gamma\subset \C^3=\{(x, y, z)\in\C^3, F_{\Gamma}(x, y, z)=0\}.
\]
The manifold ${\cal M}$ on which an ALE metric is defined is obtained by
minimally resolving the singularity at the origin of $\C^2/\Gamma$. This
desingularisation is achieved by taking ${\cal M}$ to be the zero
set of
\[
\widetilde{F}_\Gamma(x, y, z,\lambda)={F}_\Gamma(x, y, z)
+\sum_{i=1}^ra_i(\lambda)f_i(x,y,z),
\]
where $f_i$ span the ring of polynomials in $(x, y, z)$ divided by the
ideal generated by \[
<\p_x{F}_\Gamma,\p_y{F}_\Gamma, \p_z{F}_\Gamma>.
\]  
The
dimension $r$ of this ring is equal to the number of non-trivial
conjugacy classes of $\Gamma$ which is $k-1, k+1, 6, 7$ and $8$
respectively \cite{B70}.  Kronheimer \cite{K1,K2} proved that for each
$\Gamma$ a unique hyper-K\"ahler metric exists on a minimal resolution
${\cal M}$, and that this metric is precisely the ALE metric with
$\R^4/\Gamma$ as its infinity.  His construction was a combination of
the hyper-K\"ahler quotient \cite{HKLR87} with the twistor theory.

The degrees $p,q$ and $r$ are such that $\widetilde{F}_\Gamma(x, y, z,
\ll)$ is a function homogeneous of some degree $s$.
Therefore
\[
\widetilde{F}_\Gamma:\OO(p)\oplus\OO(q)\oplus\OO(r)\rightarrow \OO(s).
\]
To determine the integers $p, q, r, s$, we require that the normal
bundle to a section of ${\cal PT}\longrightarrow\CP^1$ should have the
Chern class 2. To impose this we restrict the Jacobian of the above
map to the normal bundle of the curve, and notice that the Chern class
is $p+q+r-s$ which should therefore be 2. This gives us the following
\begin{eqnarray*}
A_k & & {\T}=\{(x, y, z, \ll)\in \OO(k)\oplus\OO(k)\oplus\OO(2)
\longrightarrow \CP^1,\\
&& xy-z^k-a_1z^{k-2}-...a_{k-1}=0\},\\
D_{k-1} & & {\T}=\{(x, y, z, \ll)\in
\OO(2k)\oplus\OO(2k-2)\oplus\OO(4)\longrightarrow \CP^1,\\ 
&&x^2+y^2z+z^{k}+a_1y^2+a_2y+a_3z^{k-2}+...+a_{k}z+a_{k+1} =0\},\\
E_6&&{\T}=\{(x, y, z, \ll)\in \OO(12)\oplus\OO(8)\oplus\OO(6)
\longrightarrow \CP^1,\\
&&x^2+y^3+z^4
+y(a_1z^2+a_2z+a_3)+a_4z^2+a_5z+a_6=0\}\\
E_7&& {\T}=\{(x, y, z, \ll)\in\OO(18)\oplus\OO(12)\oplus\OO(8)\longrightarrow \CP^1,\\
&&x^2+y^3+yz^3+
y^2(a_1z+a_2)+y(a_3z+a_4)+a_5z^2+a_6z+a_7=0\}\\
E_8&&{\T}=\{(x, y, z, \ll)\in\OO(30)\oplus\OO(20)\oplus\OO(12)
\longrightarrow \CP^1,\\
&&x^2+y^3+z^5+y(a_1z^3+a_2z^2+a_3z+a_4)+a_5z^3+a_6z^2+a_7z+a_8=0\}
\end{eqnarray*}

In each case the twistor space is the three dimensional hyper-surface
$\widetilde{F}_\Gamma(x, y, z)=0$ in the rank-three bundle
$\OO(p)\oplus\OO(q)\oplus\OO(r)\rightarrow \CP^1$ where now $ x(\ll) ,
y(\ll), z(\ll) $ are coordinates up the fibres of $\OO(p), \OO(q),
\OO(r)$ respectively, $f_i=f_i(x,y,z)$, and $a_i=a_i(\ll)$ is a global
section of the appropriate power of $\OO(1)$ to make $\widetilde F$
homogeneous.  Therefore we have projections
\[
f_p:{\cal PT}\longrightarrow\OO(p),\qquad f_q:{\cal
PT}\longrightarrow\OO(q),\qquad
f_r: {\cal PT}\longrightarrow\OO(r),
\]
and we can, for example, express ${\cal PT}$ as a branched cover of
$\OO(p)\oplus \OO(q)$ branched over the singular locus of $f_p\oplus
f_q$.  Rational curves in ${\cal PT}$ project to give rational curves
in $\OO(p)\oplus \OO(q)$ tangent to the singular locus of $f_p\oplus
f_q$ and the condition that a rational curve (i.e., a section) in
$\OO(p)\oplus \OO(q)$ admits a lift to ${\cal PT}$ is that it should
be tangent to the singular locus of $f_p\oplus f_q$.  This is an
admissible subfamily by \ref{admissibility}.  Thus, in particular, we
see that the ALE spaces can be realized as admissible subfamilies of
the spaces of sections of $\OO(p)\oplus \OO(q)$, in fact in three
different ways.

\subsection{Three--dimensional Einstein--Weyl structures}
Here we consider classes of Einstein--Weyl spaces of dimension three.
Such spaces arise as the generic geometry on the space of rational
curves lying in a surface with normal bundle $\OO(2)$.  In this
section we see how examples of 3--dimensional Einstein--Weyl spaces
can be constructed by considering admissible subfamilies of relatively
simple higher dimensional families of rational curves on surfaces by
imposing intersection and tangency.

Let $\M$ be an $3$-dimensional manifold with a torsion-free connection $D$,
and a conformal structure $[h]$ which is compatible with $D$ in a sense
that $
Dh=\om\otimes h
$
for some one-form $\om$.
Here $h\in[h]$ is a representative metric in a conformal class. If
we  change this representative by $h\rightarrow \psi^2 h$,
then $\om\rightarrow \om +2\d \ln{\psi}$, where $\psi$ is a
non-vanishing function on $\M$.
The space of oriented $D$--geodesics in $\M$ is a manifold $\T$
of dimension $4$.
There exists a fixed point free
map $\tau:{\cal T}\longrightarrow {\cal T}$ which reverses an
orientation of each geodesics. Let $\gamma$ be an oriented geodesic in
$\M$, and let $U$ be a vector field tangent to $\gamma$.

The almost-complex structure on ${\cal T}$ defined by
\[
J(V)=\frac{U\times V}{\sqrt{h(U,U)}},
\]
is integrable if for any choice of $h\in[h]$
the symmetrised Ricci tensor of $D$ is proportional to $h$.
This is the conformally invariant Einstein--Weyl condition.
Hitchin \cite{Hi82} has demonstrated the
one-to-one correspondence between local solutions solutions
to the Einstein--Weyl equations, and complex surfaces
(twistor spaces) $\T$ equipped
with a fixed-point free anti-holomorphic involution $\tau$, and a
$\tau$-invariant rational curve with a normal bundle $\OO(2)$.

The EW space can be completely reconstructed form the twistor data;
Since $H^0(\CP^1, \OO(2))=\C^3$, and $H^1(\CP^1, \OO(2))=0$
we can use Kodaira's theorem \ref{Kodaira}. The EW space is a space of those
$\OO(2)$ curves which are $\tau$-invariant. The family of such curves
passing through  a given point (and its conjugate) is a geodesic of a
Weyl connection  of $D$.  To construct a conformal
structure $[h]$ consider a point on a $\tau$-
invariant $\OO(2)$ curve $C_m$. This point represents a point in a sphere
of directions $(T_m\M-0)/\R^+$, and the conformal structure on $C_m$
induces a quadratic conformal structure in $\M$.

One class of solutions can constructed  by taking an
$n$--fold covering of a neighbourhood of a $(1, n)$ curve
$\zeta=P(\ll)/Q(\ll)$ in the quadric
$\CP^1\times\CP^1$. Here $\zeta$ and $\ll$ are affine coordinates on
$\CP^1\times\CP^1$, and $P, Q$ are polynomials of degree $n$ in $\ll$.
The curve has a normal bundle $\OO(2n)$, and
the space of such curves is parametrised by $\CP^{2n+1}$ minus the
hypersurface where the resultant of $P$ and $Q$ vanishes.
In \cite{Ped} Pedersen considered an $n$-fold cover $\T$ of $\CP^1\times\CP^1$
branched along a fixed curve $\zeta=\ll^n$. The $(1, n)$ curves which meet
the fixed curve to the $n$th order give rise to curves with a normal bundle
$\OO(2)$ in $\T$ satify the condition
\[
\frac{P(\ll)}{Q(\ll)}=\ll^n-\frac{(a\ll^2+b\ll+c)^n}{Q(\ll)}.
\]
Here $a, b, c$ are complex coordinates on the resulting EW space.
More work is required to impose Euclidean or Lorentzian reality conditions.

Another class of EW spaces could be constructed by blowing up a point
on the quadric, and considering all $(1, n)$ curves passing through this
point. The resulting curves in the blown up surfaces have normal bundle
$\OO(2n-1)$. This process may also be combined with the taking the
branched covering.

The explicit forms of resulting EW structures
were determined only for $n=2$ and $n=3$ \cite{Ped,PT93}. They
 are quite
complicated but components of $h$ and $\om$ are algebraic expressions
in local coordinates $(a, b, c)$.
According to our proposal these solutions should be regarded as
`solitons' of the EW geometry. Further analysis of the corresponding
conformal invariants is required to justify this claim.

Yet another class of examples arises form a rational curve in $\CP^2$
of degree $d>1$ whose singularities are $D=(d-1)(d-2)/2$ distinct nodes.
Let $\T$ be a surface obtained from $\CP^2$ by blowing up points on
$S\cup N$, where $S$ is a set of $s$ non-singular points  and $N$ is the
set of $n$ nodes on the curve. The resulting curve in the blown up space
will have a normal bundle $\OO(k)$ where $k$ depends on $n$ and $s$.

\subsection{Generalised W\"unschmann conditions}
As a final application of the above ideas, related to the previous
subsection, we consider an $n$-dimensional family of curves in a
surface $\Sigma$.  Such curves have a natural lift to the bundle of
$(n-2)$--jets of such curves $\T=J^{n-2}\Sigma$ over $\Sigma$.  This
family of curves will be expressed as a family of solutions to an
ordinary differential equation of order $n$ below and will be
considered as a family of curves in $\T$.  We then ask whether we can
characterise those differential equations that give rise to an
admissible family of rational curves in $\T$.  This turns out to be
given by what we will refer to as the {\em W\"unschmann condition},
W\"unschmann (1905), and its generalisations on the coefficients of
the equations.  One can ask whether the projection of this family to a
families of curves in $J^r\Sigma$, $r<n-2$, are admissible, and these
will lead to further generalisations of the W\"unschmann conditions.

Consider a relation of the form
\[
\Psi(x, y, m)=0
\]
between the complex variables $m= (m_1, m_2, ...m_n)$ (local coordinates
on an $n$-dimensional manifold $\M$), and $(x, y)$
(complex local coordinates on a two-dimensional  manifold ${\Sigma})$.
For each fixed choice of $(x, y)$  the relation
defines an $\alpha$-surface in  $\M$. Conversely each choice of $m$ defines
a curve $C_m$ in $\Sigma$. We can  apply the
implicit function theorem to
$\Psi=0$, and regard $C_m$ as a graph
$
x\longrightarrow (x, y=Z(x, m)).
$
Consider a system of equations consisting of
$y=Z(x, m)$, and the first $(n-1)$ derivatives with respect to $x$.
Solving this system for $m$, and differentiating once more with respect
to $x$ yields the ODE
\[
y^{(n)}:=\frac{\d^n y}{\d x^n}=F(x, y, y', ..., y^{(n-1)}),
\]
where the explicit form of $F$ is completely determined by $\Psi$.

Asking that the $\alpha$-sufaces in $\M$ arise from specific geometric
structures on $\M$ (which from now on will be identified with the
space of solutions to the ODE) imposes additional constraints on $F$.
This idea goes back to Cartan \cite{C41}, and his program of
`geometrising' ODEs.

A different approach based on twistor theory was suggested by Hitchin
\cite{H82} and LeBrun \cite{L80}.  In this approach the relation
$\Psi=0$ represents part of a rational curve in $\Sigma=\T$ with a prescribed
normal bundle. The local differential geometry of $\M$ is encoded in
the global complex structure of $\T$, and the globality of the curve
implies that $\alpha$-curves are the geodesics of a projective
structure.  The ODE does not explicitly appear in the correspondence
between $\M$ and $\T$.  The details of the Hitchin-LeBrun construction and
its connection with the ODE approach have only been worked out
fully for $n=2$.  In this case there exists an embedding of rational
curve with a normal bundle $\OO(1)$ in $\T$ if and only if
\[
\frac{\d^2}{\d x^2}F_{11}-4\frac{\d}{\d x}F_{01}-F_1\frac{\d}{\d x}F_{11}
+4F_1F_{01}-3F_0F_{11}+6F_{00}=0,
\]
where $F_0=\p F/\p y, F_1=\p F/\p y'$ and $\d/\d x=\p/\p x+y'\p/\p y
+F\p/\p y'$.
The two-dimensional moduli
space $\M$ of $\OO(1)$ curves (the space of solutions to the ODE is
in this case equipped with a projective structure, in the sense that
the $\alpha$-surfaces (here curves) of constant $(x, y)$
are the geodesics of a torsion-free projective connection. Conversely, given a
projective structure on $\M$ one defines ${\T}$ as a quotient space
of the foliation of  $P(T\M)$ by the orbits of the geodesic flow.
Each projective tangent space $P(T_m\M)$ maps to a rational
curve with self-intersection number one in ${\T}$.

The case $n=3$ which goes back to Cartan \cite{C41} and  was
recently revisited by Tod \cite{T00}.
The conformal structure on $\M$ is defined by demanding
that hyper-surfaces $z\subset \M$ corresponding to points in $\T$
are null (it could also be defined by declaring the curves in ${\cal
  M}$ that correspond to points of ${\cal T}J^1\Sm$ to be null
geodesics).
 This conformal structure doesn't depend on $(x, y)\in\Sm$
if $F(x, y, y', y'')$ satisfies a second-order differential
constraint
\[
\frac{1}{3}F_{{2}}{\frac {\d}{\d x}}F_{{2}}-
\frac{1}{6}{\frac {\d^{2}}{\d{x}^{2}}
}F_{{2}}+\frac{1}{2}{\frac {\d}{\d x}}F_{{1}}-{\frac {2}{27}}\left (F_{
{2}}\right )^{3}-\frac{1}{3}F_{{2}}F_{{1}}-F_{{0}}=0.
\]
This constraint has already appeared in a work
W\"unschmann \cite{W05}.

 The only other case which has attracted some attention is $n=4$.
Bryant \cite{B91} has shown that there exist a correspondence between
a class of fourth order ODEs, and exotic non-metric holonomies in dimension
four. The conditions on $F$ are only implicit in Bryant's work.

We shall say that $F$ satisfies the generalised W\"unschmann conditions if
there exists
$SL(2, \C)$ invariant paraconformal structure
\[
T\M\cong \C^2\odot \C^2\odot ... \odot \C^2=\mbox{S}^{n-1}{\C^2}.
\]
The explicit form of the generalised W\"unschmann conditions has been
worked out recursively in \cite{DT03}. For example if $n=4$ one gets
\begin{eqnarray*}
&&{\frac {11}{1600}}\,\left (F_{{3}}\right )^{4}-{\frac {9}{50}}\,
\left (F_{{3}}\right )^{2}{\frac {\d}{\d x}}F_{{3}}-{\frac {1}{200}
}\,\left (F_{{3}}\right )^{2}F_{{2}}+{\frac {21}{100}}\,\left ({
\frac {\d}{\d x}}F_{{3}}\right )^{2}+{\frac {1}{50}}\left ({\frac {\d
}{\d x}}F_{{3}}\right )F_{{2}}\\
&&-{\frac {9}{100}}\,\left (F_{{2}}
\right )^{2}+{\frac {7}{20}}\,F_{{3}}(x){\frac {\d^{2}}{\d{x}^{2}}}F_{{3
}}-\frac{1}{5}\,{\frac {\d^{3}}{\d{x}^{3}}}F_{{3}}+\frac{3}{10}\,{\frac {\d^{2}}{\d{
x}^{2}}}F_{{2}}-\frac{1}{4}\,F_{{3}}{\frac {\d}{\d x}}F_{{2}}-F_{{0}}=0,\\
&&
\frac{9}{4}\,F_{{3}}{\frac {\d}{\d x}}F_{{3}}-\frac{3}{2}\,{\frac {\d^{2}}{\d{x}^{2}}
}F_{{3}}+3\,{\frac {\d}{\d x}}F_{{2}}-\frac{3}{8}\,\left (F_{{3}}\right
)^{3}-\frac{3}{2}\,F_{{2}}F_{{3}}-3\,F_{{1}}=0.
\end{eqnarray*}
If $F$ satisfies these conditions, then the space ${\cal M}$ of solutions 
to the corresponding ODE is equipped with a torsion--free connection with
holonomy $G_3$ in the terminology of \cite{B91}. 

In general $F$ has to satisfy an over-determined system of $n-2$ PDEs, and
a priori it is not clear that any solutions exist. It can however
be verified that the method of admissible curves provides (some) solutions
to all the constraints. For example one can consider the blow-ups applied to
described in the last section to find a simple solutions
$F=(4/3)(y''')^2/y''$, or $F=(ay'''+b)^{4/3}$ when $n=4$.


\appendix
\section{Appendix: Rational curves and their embedings}

\def\theequation{\thesection{A}\arabic{equation}}
Let $\C^2$ be  a vector space
with coordinates $\pi=(\pi_0, \pi_1)$.
Remove
$\pi=(0, 0)$ and use $[\pi]$ as homogeneous coordinates on $\CP^1$.
We shall also use the affine coordinate $\ll=\pi_{0}/\pi_{1}$.
Holomorphic functions on $\C^2-0$ extend to holomorphic functions on $\C^2$
(Hartog's Theorem). Therefore homogeneous functions on $\CP^1$ are polynomials.
In particular, holomorphic functions homogeneous of degree $0$
are constant (Liouville theorem).
Let us summarize some facts about holomolrphic line bundles
over $\CP^1$. First define a tautological line bundle
\[
\OO(-1)=\{(\ll, (\pi_0, \pi_1))\in \CP^1\times\C^2|\ll=\pi_0/\pi_1\}.
\]
Other line bundles can be obtained from $\OO(-1)$ by algebraic operations:
\[
\OO(-n)=\OO(-1)^{\otimes n}, \qquad \OO(n)=\OO(-n)^*, \qquad\OO=\OO(-1)\otimes\OO(1),
\qquad n\in \NN.
\]
Equivalently  ${\cal O}(n)$ denotes the line bundle over $\CP^1$ with transition functions
$\ll^{-n}$ from the set $\ll\neq\infty$ to $\ll\neq 0$ (i.e.\ Chern class
$n$).  Its sections
are given by functions homogeneous of degree $n$ in a sense that
$
f(\xi\pi)=\xi^nf(\pi).
$
These are polynomials in $\ll$ of degree $n$ with complex
coefficients.
The theorem of Grothendick states that all holmorphic line bundles
over a rational curve are equivalent to $\OO(n)$ for some $n$.
The spaces of global  sections, and the first cohomology groups are
\[
 H^0(\CP^1,{\cal O}(n))=\left\{ \begin{array}{ll}
                       0    & \mbox{ for}\; n<0 \\
                       \C^{n+1} & \mbox{for}\; n\geq 0.
                       \end{array}
                           \right.\qquad
\label{dimh1}
 H^1(\CP^1,{\cal O}(-n))=\left\{ \begin{array}{ll}
                       0    & \mbox{ for}\; n<2 \\
                       \C^{n-1} & \mbox{ for}\; n\geq 2 .
                       \end{array}
                           \right.
\]

The following result of Kodaira underlies the twistor approach to
curved geometries. Let ${\T}$ be a complex manifold of dimension $d+r$.
A pair $(E, {\M})$ is called {\it a complete analytic family
of compact sub-manifolds  of ${\T}$ of dimension $d$} if
\begin{itemize}
\item
$E$ is a complex analytic sub-manifold of ${\T}\times{\cal M}$
of codimension $r$ with the property that for each $m\in {\cal M}$
the intersection $C_m:=E\cap ({\T}\times m)$ is a compact
sub-manifold of ${\T}\times m$ of dimension $d$.
\item
There exists an isomorphism
\[
T_m{\cal M}\simeq H^0(C_m, NC_m)
\]
where $NC_m\longrightarrow C_m$ is the normal bundle of
$C_m$ in  ${\T}$.
\end{itemize}
\begin{theo}[Kodaira\cite{Ko63}]
Let $E$ be a complex compact sub-manifold of ${\T}$ of dimension
$d$, and let $NE$ be the normal bundle of $E$ in ${\T}$.
If $H^1(E, NE)=0$ then there exists a complete analytic family
$(E, {\cal M})$ such that $E=E(m_0)$ for some $m_0\in{\cal M}$.
\label{Kodaira}
\end{theo}
We will apply the above theorem to the situation when
${\T}$ is a  twistor space and $E=\CP^1$. Roughly speaking,
the moduli space ${\cal M}$ is the `arena' of differential geometry
and integrable systems.


\begin{thebibliography}{jafsdl}


\frenchspacing

\bibitem{AW77} Atiyah, M.F.\ \& Ward, R.S.\ (1977) Instantons and
  algebraic geometry, {\em Comm. Math. Phys.}, {\bf 55}, 111-124. 

\bibitem{BG03} Bernstein, J.\& Gindikin S.(2003) Notes on Integral
geometry for Manifolds of Curves, Amer.Math Soc.Transl.(2) {\bf
vol 210}, 57--80

\bibitem{B70} Brieskorn, E. (1970) Singular elements of semisimple
algebraic groups, Actes Congres Intern. Math. {\bf vol 2}, 279-284.


\bibitem{B91} Bryant, R.L. (1991) Two exotic holonomies in dimension four,
path geometries, and twistor theory.
Proc. Symp. Pure. Maths. Vol. 53, 33--88



\bibitem{C41} Cartan, E.(1941) La Geometria de las Ecuaciones
  Diferenciales de Tercer Orden. Rev. Mat. Hispano-Amer. 4, 1--31

\bibitem{DM00} Dunajski, M.\& Mason, L.J. (2000)
Hyper-K\"ahler Hierarchies and their twistor theory
Comm. Math. Phys {\bf 213} 641-672

\bibitem{DM03} Dunajski, M. \& Mason, L.J. (2003)
{\em Twistor theory of hyper-K{\"a}hler metrics with hidden symmetries},
 J. Math. Phys. {\bf 44}, 3430-3454.

\bibitem{DT03} Dunajski, M. \& Tod K.P. (2004) A geometry of $n$th order ODEs.
Preprint.


\bibitem{G82} Gindikin, S. (1982) Bundles of differential forms and
the Einstein equation, Nuclear Phys. {\bf 36}, 2(8), 537-548.

\bibitem {G84} Gindikin S.(1984) Reduction of manifolds of
rational curves and related problems of differential equations,
Funktionsal'nyi Analiz Ego Prilozheniya, {\bf 18} No.4, 14--39
(Russian).  English  translation.Functional Anal.Appl.{\bf
18},4,278--298

\bibitem{G85}
Gindikin, S. (1985)  Some solutions of the self-duality
Einsteinequations, Funktionsal'nyi Analiz Ego Prilozheniya, Vol.
19, No.3, pp58-60 (Russian).  English  translation English
translation.Functional Anal.Appl.{\bf 19},3,210--213

\bibitem{G90} Gindikin, S. (1990) Generalized conformal structures,
in {\em Twistors in Mathematics and Physics}, eds Bailey, T.N. \&
Baston, R.J., LMS lecture notes series 156, CUP.

\bibitem{Hi82} Hitchin, N. (1982) Monopoles and geodesics,
  Commun. Math. Phys. {\bf 83}, 589-602.


\bibitem{H82} Hitchin, N.\ (1982) Complex manifolds and Einstein's
equations, in {\em Twistor Geometry and Non-Linear systems}, Springer
LNM 970,
ed. Doebner, H.D.\ \& Palev, T.D..

\bibitem{HKLR87} Hitchin, N.J., Karlhede, A., Lindstrom, U.\& Rocek,
M. (1987) Comm. Math. Phys {\bf 108} 21


\bibitem{Ko63} Kodaira, K. (1963)
On stability of compact submanifolds of complex manifolds,
 Am. J. Math.  {\bf 85}, 79-94.

\bibitem{K1} Kronheimer, P. (1989)
 The Construction Of ALE Spaces As Hyperkahler Quotient,
J.Diff.Geom.  {\bf 29}  665



\bibitem{K2} Kronheimer, P. (1989)
A Torelli Type Theorem For Gravitational Instantons,
J.Diff.Geom.  {\bf 29}  685

\bibitem{L80} LeBrun, C.R. (1980) Spaces of complex geodesics and
  related structures, D.Phil. thesis, Oxford University,


\bibitem{MS92} Mason, L.J. \& Sparling, G.A.J. (1992) Twistor correspondences
for the soliton hierarchies, J. Geom. Phys., {\bf 8}, 243-271.


\bibitem{MW96}
Mason, L.J., and Woodhouse, N.M.J. (1996) Integrability, Self-duality
and Twistor theory, LMS Monograph, CUP.

\bibitem{Ped} Pedersen, H. (1986)
Einstein--Weyl spaces and $(1, n)$ curves in the quadric surface,
Ann.Global Anal.Geom. 4, No 1, 89--120.



\bibitem{PT93} Pedersen, H. \& Tod, K.P. (1993) Three-Dimensional Einstein--Weyl Geometry,
Adv. in Math. {\bf 97} 74-109.


\bibitem{Pe76} Penrose, R. (1976) Nonlinear gravitons and curved
twistor theory, Gen. Rel. Grav. {\bf 7}, 31-52.

\bibitem{T00} Tod, K.P. (2000)
Einstein-Weyl Spaces and Third Order Differential Equations,
J. Math, Phys. 41, 5572.




\bibitem{Wa77} Ward, R.S. (1977) On self-dual gauge fields, Phys.
Lett. {\bf 61A}, 81-82.

\bibitem{Wa81}  Ward, R.S. (1981) Ansatze for self-dual Yang-Mills
  fields, Comm. Math. Phys., {\bf 80}, 563-74.

\bibitem{W05} W\"unschmann, K.W. (1905) \"Uber Beruhrungsbedingungen
  bei Dierentialgleichungen, Dissertation, Greifswald.

\end{thebibliography}
\end{document}